\documentclass[12pt,
 reprint,
 amsmath,amssymb,superscriptaddress,
 aps,
 prl,
groupedaddress
]{revtex4-1}

\usepackage{graphicx}
\usepackage{dcolumn}
\usepackage{bm}
\usepackage{color}
\usepackage{ulem}
\usepackage{float}
\usepackage{textcomp}
\restylefloat{table}
\usepackage{verbatim}
\usepackage{epstopdf}
\usepackage[mediumspace,mediumqspace,Grey,squaren]{SIunits}

\newcommand{\RN}[1]{%
  \textup{\uppercase\expandafter{\romannumeral#1}}%
}

\begin{document}

\title{Realization of a Valley Superlattice}
\date{\today}

\author{M. A.\ Mueed}
\author{Md. Shafayat\ Hossain}
\author{I.\ Jo}
\author{L. N.\ Pfeiffer}
\author{K. W.\ West}
\author{K. W.\ Baldwin}
\author{M.\ Shayegan}
\affiliation{Department of Electrical Engineering, Princeton University, Princeton, New Jersey 08544, USA}

\begin{abstract}

In a number of widely-studied materials, such as Si, AlAs, Bi, graphene, MoS$_2$, and many transition metal dichalcogenide monolayers, electrons acquire an additional, spin-like degree of freedom at the degenerate conduction band mimina, also known as ``valleys". 
External symmetry breaking fields such as mechanical strain, or electric or magnetic fields, can tune the valley-polarization of these materials 
making them suitable candidates for ``valleytronics". 
Here we study a quantum well of AlAs, where the two-dimensional electrons reside in two energetically degenerate valleys. 
By fabricating a strain-inducing grating on the sample surface, we engineer a spatial modulation of the electron population in different valleys, i.e., a ``valley superlattice" in the quantum well plane. 
Our results establish a novel manipulation technique of the valley degree of freedom, paving the way to realizing a valley-selective layered structure in multi-valley materials, with potential  application in valleytronics.
\end{abstract} 

\maketitle
 
With the ubiquitous, Si-based electronics approaching its fundamental physical limitations, there is an active push to explore novel device concepts. An emergent field is ``valleytronics" which harnesses electrons' valley degree of freedom instead of the conventional charge-based operations \cite{Rycerz.NatPhy.2007,Schaibley.NatMat.2016,Shayegan.Physica.2006,Gunawan.PRL.2006,Gunawan.PRB.2006}. 
Valleys are energetically-degenerate pockets in the momentum space of crystals that possess multiple conduction band minima at equal energies \cite{Rasolt.Book,Rasolt.PRL.1986,Shkolnikov.PRL.2004,Shkolnikov.PRL.2005,Gunawan.NatPhy.2007,Feldman.Science.2016,Zhu.NatComm.2017,Schaibley.NatMat.2016,Shayegan.Physica.2006,Gunawan.PRL.2006,Gunawan.PRB.2006,Rycerz.NatPhy.2007,Young.NatPhy.2012,Zeng.NatNano.2012,Jones.NatNano.2013,Gokmen.PRB.2008}. It is possible to lift the valley-degeneracy by straining or placing the materials under electric or magnetic fields \cite{Rasolt.Book,Rasolt.PRL.1986,Shkolnikov.PRL.2004,Shkolnikov.PRL.2005,Gunawan.NatPhy.2007,Feldman.Science.2016,Zhu.NatComm.2017,Schaibley.NatMat.2016,Shayegan.Physica.2006,Gunawan.PRL.2006,Gunawan.PRB.2006,Rycerz.NatPhy.2007,Young.NatPhy.2012,Zeng.NatNano.2012,Jones.NatNano.2013,Gokmen.PRB.2008}. The resulting polarizability makes valleys analogous to electron's spin degree of freedom \cite{Rasolt.Book,Rasolt.PRL.1986,Zeng.NatNano.2012,Jones.NatNano.2013,Shayegan.Physica.2006,Shkolnikov.PRL.2004,Shkolnikov.PRL.2005,Gunawan.PRL.2006,Gunawan.PRB.2006,Gunawan.NatPhy.2007,Rycerz.NatPhy.2007,Schaibley.NatMat.2016}, which is at the heart of ``spintronics'', another branch of unconventional, next-generation electronics 
\cite{Zutic.RMP.2004,Jansen.NatMat.2012}. Considering the similarities, valleytronics also offers an intriguing parallel route for novelty. For example, digital information may be stored and processed by compelling electrons to selectively occupy one valley or another. Moreover, coupling the valley degree of freedom with polarized light can result in exotic opto-electronic properties \cite{Schaibley.NatMat.2016,Jones.NatNano.2013,Zeng.NatNano.2012}.


\begin{figure*}
\includegraphics[width=.85\textwidth]{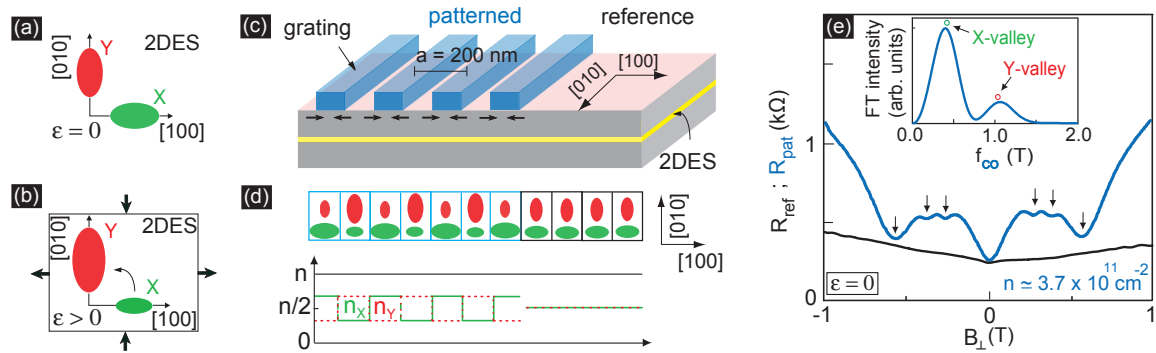}
\caption{\label{fig:Fig1} 
(a) For 2D electrons confined to a wide AlAs QW, the degenerate X- and Y-valleys are equally occupied. (b) Positive symmetry-breaking, uniaxial strain ($\varepsilon=\varepsilon_{[100]}-\varepsilon_{[010]}$) in the QW plane transfers 2D electrons from the X-valley to Y-valley, and vice versa for negative $\varepsilon$. 
(c) Sample schematic. The surface is partially covered by a grating made of strips of electron-beam resist (blue strips). Each box in (d) represents the valley occupation of the 2DES directly above it, as also indicated by the 
density plots at the bottom of the panel. Here we portray $n_X$ and $n_Y$ plots as step functions for simplicity; however, a sinusoidal variation may be more realistic. Note that the total density \textit{n} is constant in all boxes. (e) Magneto-resistance traces for the patterned region (blue) and reference region (black) for $n\simeq3.7\times10^{11}$ cm$^{-2}$. The inset shows the Fourier transform (FT) spectrum of the oscillations (marked by vertical arrows) in the blue trace; the green and red circles mark the expected positions of the maxima in the FT spectrum.
}
\end{figure*}

In this study, we investigate a two-dimensional electron system (2DES) contained in a 12-nm-wide AlAs quantum well (QW) structure. The QW, located
143 nm below the surface, is sandwiched between two AlGaAs spacer layers, all grown via molecular beam epitaxy on a (001) GaAs substrate (see Supplemental Material for details). The high transport mobility ($\sim3\times10^5$ cm$^2/$Vs for our samples) makes the AlAs 2DES well-suited for potential valleytronics applications. Its valley degree of freedom arises from a two-fold valley-degeneracy \cite{Shayegan.Physica.2006}. We denote these valleys as X and Y with the major axes lying along [100] and [010], respectively (see Fig. 1(a)). Without any symmetry-breaking, uniaxial strain ($\varepsilon$) applied to the QW plane, electrons are equally distributed between the two valleys, where they possess an anisotropic Fermi surface with longitudinal and transverse effective masses of $m_l = 1.05$ and $m_t = 0.205$, in units of the free electron mass. Under finite $\varepsilon$, however, electrons move from one valley to the other, as described in Fig. 1(b). Here $\varepsilon=\varepsilon_{[100]}-\varepsilon_{[010]}$, where $\varepsilon_{[100]}$ and $\varepsilon_{[010]}$ are the strain
values along [100] and [010]. (The single-particle valley splitting is given by $\varepsilon E_2$ where $E_2$ is the deformation potential, $\simeq5.8$ eV for AlAs \cite{Shayegan.Physica.2006}.)  
Taking advantage of this inter-valley transfer of electrons, we report here how to engineer a 2D valley superlattice, namely, to reconstruct a valley-degenerate 2DES plane into multiple strips where the X- and Y-valley alternate as the majority-valley species. 
We demonstrate such a lateral modulation of the X- and Y-valley occupation through measurements of commensurability oscillations. It is worth emphasizing at the outset that, unlike the typical commensurability phenomena \cite{Weiss.Europhys.1989, Gerhardts.PRL.1989, Winkler.PRL.1989,Beenakkaer.PRL.1989}, the total charge density in our system stays uniform, and it is the modulation of individual valley densities that leads to the commensurability oscillations.

Figure 1(c) illustrates our approach to realizing a valley superlattice. We partially pattern the surface of a standard Hall bar sample along [100] with a grating (shown as blue strips) of 200-nm periodicity, made of negative electron-beam resist. When cooled to low temperatures, strain develops at the interface of each resist strip and the sample surface, thanks to their different thermal contraction coefficients \cite{Skuras.APL.1997,Long.PRB.1999,Davies.PRB.1994}. The strain field can then couple to the AlAs QW through the deformation potential. Now, assuming the 2DES below each resist strip is under negative $\varepsilon$, it must be under positive $\varepsilon$ between two strips. The periodic grating thus should subject the 2DES to varying $\varepsilon$ of the same periodicity, locally breaking the valley degeneracy. As a result, the 2DES (with total density $n$) should separate into multiple 200-nm-wide regions, each of which is partially valley-polarized with the majority electron population periodically alternating between the X- and Y-valley (of density $n_X$ and $n_Y$). Such a density profile is depicted in the blue boxes of Fig. 1(d), each corresponding to the 2DES section directly above it, as well as the $n_X$ and $n_Y$ plots. While we expect $n_X>n_Y$ below each resist strip and $n_Y>n_X$ in between, the total density $n=n_X+n_Y$ is independent of strain \cite{Shayegan.Physica.2006} and should stay uniform for the entire patterned region. Note that AlAs and GaAs are generally piezo-electric. However, they belong to the symmetry group \textit{43m} and are \textit{not} piezo-electric along [100] \cite{Nye.Book,Skuras.APL.1997}, the direction of the surface grating in our sample. This rules out any electric field modulation due to the periodic surface-strain along [100], implying a uniform total 2DES density in that direction \cite{Nye.Book,Skuras.APL.1997}, which is in agreement with the constant total density profile depicted in Fig. 1(d). For the unpatterned (reference) region, the 2DES remains valley-degenerate ($n_X=n_Y=n/2$).

To establish the existence of a valley superlattice in the patterned region, one needs to probe the spatial valley densities, for which we employ the commensurability (also known as Weiss) oscillations (COs) technique \cite{Weiss.Europhys.1989, Gerhardts.PRL.1989, Winkler.PRL.1989,Beenakkaer.PRL.1989}. By passing current along the grating direction ([100]) under perpendicular magnetic field ($B_{\perp}$), we look for low-field magneto-resistance minima, expected whenever the cyclotron orbit diameter ($2R_{C}$) of electrons becomes commensurate with the period ($a$) of the 2DES density modulation. The exact condition for commensurability is $2R_C/a= i-1/4$, where $i$ is an integer \cite{Weiss.Europhys.1989,Beenakkaer.PRL.1989,Gerhardts.PRL.1989,Winkler.PRL.1989}. The $B_{\perp}$-positions of the COs minima depend on the 2DES density according to $2R_C=2k_{F}/eB_{\perp}$, where $k_F=\sqrt{2\pi n}$ for a single-valley, isotropic 2DES. Before discussing the COs for the bi-valley, anisotropic AlAs 2DES, we emphasize an important point. The prerequisite for COs is a periodic density modulation, usually rendered in the \textit{total} density by various means such as an optical interference pattern \cite{Weiss.Europhys.1989}, electrostatic gating \cite{Winkler.PRL.1989}, or strain-inducing surface superlattice in piezo-electric materials (along the appropriate crystal direction) \cite{Skuras.APL.1997}. 
In stark contrast to the typical COs scenario, the \textit{total} density in our samples should be uniform for strain modulation along [100], thanks to the underlying crystal symmetry mentioned previously. This compelling difference guarantees that COs we report here, as discussed below, must originate from the modulation in the individual valley densities (see Fig. 1(d)), i.e., a \textit{valley superlattice}. 

\begin{figure*}
\includegraphics[width=1\textwidth]{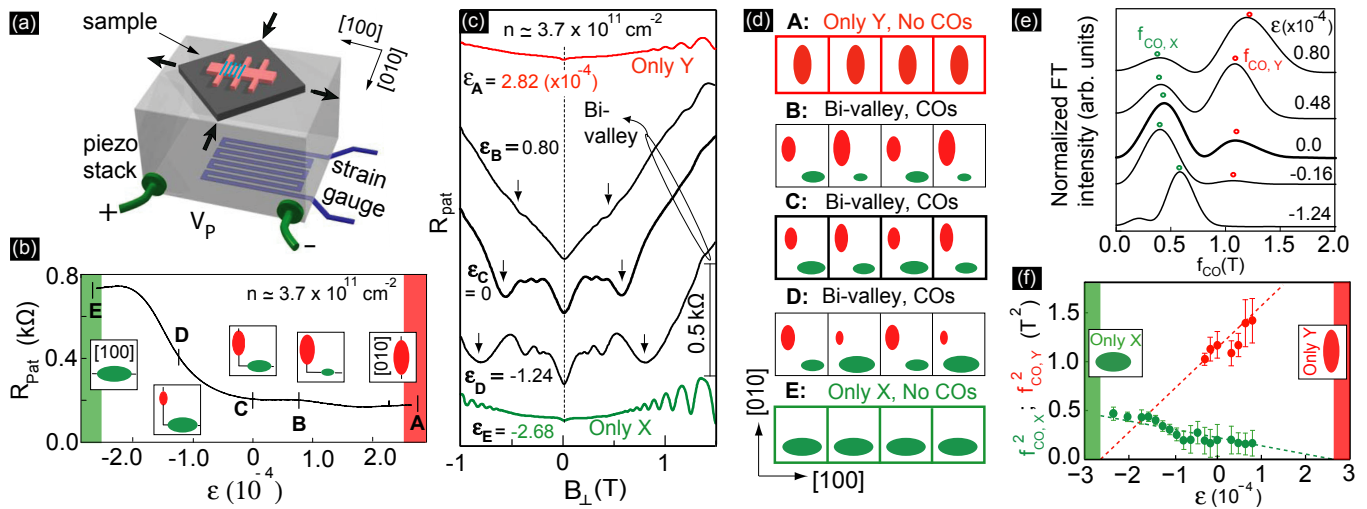}
\caption{\label{fig:Fig1} (a) Schematic of the experimental setup for the application of global, uniaxial strain ($\varepsilon$) with the sample and a strain gauge glued to the opposite faces of a piezo-actuator. (b) Piezo-resistance of the patterned region as a function of $\varepsilon$ for $n\simeq3.7\times10^{11}$ cm$^{-2}$. 
We mark points \textbf{A} - \textbf{E} on the trace. (c) Magneto-resistance traces corresponding to points \textbf{A} - \textbf{E} of (b). The traces are vertically offset for clarity. Arrows mark the prominent COs minima observed in the black traces. At higher fields Subhnikov-de Hass oscillations, originating from the formation of well-defined Landau levels, are observed. (d) Schematic of the spatial density profile of the X- and Y-valley for points \textbf{A} - \textbf{E}. Note that the density distributions shown are only qualitative. (e) Examples of Fourier transform (FT) spectra for different strain values. (f) Plot of the observed FT frequencies squared versus $\varepsilon$. The dashed green and red lines represent the expected $\varepsilon$-induced evolution of the CO frequencies from the X- and Y-valleys for $n\simeq3.7\times10^{11}$ cm$^{-2}$.
}
\end{figure*}

Our transport measurements are carried out in a $^3$He cryostat at a temperature of 0.3 K. 
By varying the illumination time at low temperatures, we tuned $n$ between 2.0 to
$3.7\times10^{11}$ cm$^{-2}$. Figure 1(e) shows the magneto-resistance from the patterned (blue trace, $R_{pat}$) and reference (black trace, $R_{ref}$) regions for $n\simeq3.7\times10^{11}$ cm$^{-2}$, as current is passed along [100]. 
Near $B_{\perp}=0$, $R_{pat}$ exhibits a pronounced V-shaped resistance minimum followed by multiple minima (vertical arrows) at slightly higher $B_{\perp}$. The absence of such minima in $R_{ref}$ points to their COs' origin. To further corroborate, we also present in Fig. 1(e) inset the Fourier transform (FT) spectrum of the oscillations observed in $R_{pat}$. Note that, for an AlAs 2DES, there are two relevant $k_F$ for COs along the modulation direction of [100], each for the elliptical Fermi surfaces of the X- and Y-valley, $k_{F,X}$ and $k_{F,Y}$, respectively; here $k_{F,X}^{2}=2\pi n_X\sqrt{(m_t/m_l)}$ and $k_{F,Y}^{2}=2\pi n_Y\sqrt{(m_l/m_t)}$ \cite{Gunawan.PRL.2004}. Based on the expression $f_{CO}=2\hbar k_{F}/ea$, and assuming the \textit{average} densities of $n_X=n_Y=n/2$, we
mark with green and red circles the expected COs frequencies for the X- and Y-valleys, respectively. 
The peaks in the FT spectrum closely agree with the expected COs' frequencies, justifying our $n_X=n_Y=n/2$ assumption. This establishes that the oscillations in $R_{pat}$ are indeed COs, providing strong evidence for the periodic modulation in both the X- and Y-valley density in the patterned region. 

Next we present data in Fig. 2 to further consolidate
the existence of the valley superlattice as we manipulate
the individual valley densities by applying global, uniaxial
strain. For this purpose, we glue the back side of our sample, with its length along [100], to one side of a stacked piezo-electric actuator, and a strain gauge to the opposite side (see Fig. 2(a)). Via applying a voltage ($V_{P}$) to the actuator's leads, a global, in-plane $\varepsilon$ ($=\varepsilon_{[100]}-\varepsilon_{[010]}$) is added to the 2DES. 
First we address the strain-induced piezo-resistance behavior of the patterned region for $n\simeq3.7\times10^{11}$ cm$^{-2}$, taken at $B_{\perp}=0$. The resistance profile, shown in Fig. 2(b), is typical of the bi-valley AlAs 2DES \cite{Shayegan.Physica.2006} and reflects the inter-valley transfer of electrons and their anisotropic mobility in the X- and Y-valley. At large negative $\varepsilon$, the resistance saturates when only the X-valley, which has a large effective mass ($m_l$) and therefore high resistance along [100], is occupied (point \textbf{E}). For increasing $\varepsilon$, however, the resistance starts to drop as electrons begin the inter-valley transfer (points \textbf{D} - \textbf{B}) and eventually saturates at large positive $\varepsilon$. Now all electrons reside in the Y-valley which has low resistance along [100] (point \textbf{A}). The reduction in resistance stems from the Y-valley electrons' smaller effective mass ($m_t$) and higher mobility along the current direction ([100]).  
We remark that the saturating behavior of resistance at large $|\varepsilon|$ is consistent with the green and red shaded regions, expected for the only X- or only Y-valley occupation in the 2DES, according to previous measurements \cite{Gokmen.PRB.2008,Gokmen.Natphy.2010}.

In Fig. 2(c), we show a series of low-$B_{\perp}$ magneto-resistance traces corresponding to points \textbf{A} - \textbf{E} of Fig. 2(b). At $\varepsilon=0$ (point \textbf{C}), $R_{pat}$ manifests pronounced COs minima, as previously addressed in Fig. 1(e). For moderate negative $\varepsilon$ (point \textbf{D}), the minima (marked by vertical arrows) move out to higher $B_{\perp}$, expected for the growing X-valley population which increases $k_{F,X}$. However, for point \textbf{E}, when the 2DES is completely valley-polarized (Only X) at large $\varepsilon<0$, the COs disappear and instead strong Subhnikov-de Hass oscillations are seen. 
Much like the $\varepsilon<0$ case, the COs are present (vertical arrows) only at moderate positive $\varepsilon$ but not at large $\varepsilon>0$ (point \textbf{A}) when the 2DES is once again fully valley-polarized (Only Y) (see the top trace of Fig. 2(c)). Note that if any residual piezo-electric effect were present along [100], it should modulate the total density even in the fully valley-polarized 2DES and give rise to COs. The absence of COs for the only X or Y cases therefore unambiguously rules out the existence of piezo-electric effect along [100], as expected \cite{Nye.Book,Skuras.APL.1997}. This also reaffirms that the observed COs are indeed made possible by the valley superlattice, which exists only in the bi-valley cases, i.e., when both the X- and Y-valleys are occupied (points \textbf{B} - \textbf{D}). Figure 2(d) further elucidates how the valley modulation (present in the bi-valley cases) from the local strain field evolves under increasing global $\varepsilon$ from the piezo-actuator and is eventually nullified when large $|{\varepsilon}|$ fully valley-polarizes the entire 2DES. 

Figures 2(e) and 2(f) present a more quantitative picture of how each valley contributes to the COs as a function of $\varepsilon$. In Fig. 2(e) we show FT spectra at few representative strains. We plot these COs frequencies squared as a function of $\varepsilon$ in Fig. 2(f). The dashed green and red lines are based on the expressions, $f_{CO,Y}^{2}=h^{2}\sqrt{m_l/m_t}{(n+\Delta n)}/\pi e^{2}a^{2}$ and $f_{CO,X}^{2}=h^{2}\sqrt{m_t/m_l}{(n-\Delta n)}/\pi e^{2}a^{2}$, where $f_{CO,Y}$ and $f_{CO,X}$ are the expected COs frequencies for the Y- and X- valleys at a given $\Delta n$ (= $n_Y-n_X$), the density difference between the two valleys \cite{Gunawan.PRL.2004}. Here we convert $\Delta n$ to $\varepsilon$ according to their empirical relation for $n\simeq3.7\times10^{11}$ cm$^{-2}$, as reported in Ref. \cite{Gokmen1.PRB.2008}, and the dashed lines are based on the relation $f_{CO}=2\hbar k_F/ea$ \cite{Gunawan.PRL.2004}. As more electrons are transferred from the X- to the Y-valley with increasing $\varepsilon$ (see Fig. 2(d)), $k_{F,X}$ shrinks and thereby $f_{CO,X}$ decreases. In contrast, $f_{CO,Y}$ increases since $k_{F,Y}$ gets enhanced. As seen in Fig. 2(f), the COs' frequencies nicely agree with the dashed lines, confirming that they originate from the valley modulation of the X- and Y- valleys. (See Supplemental Material for more details).

\begin{figure}
\includegraphics[width=.27\textwidth]{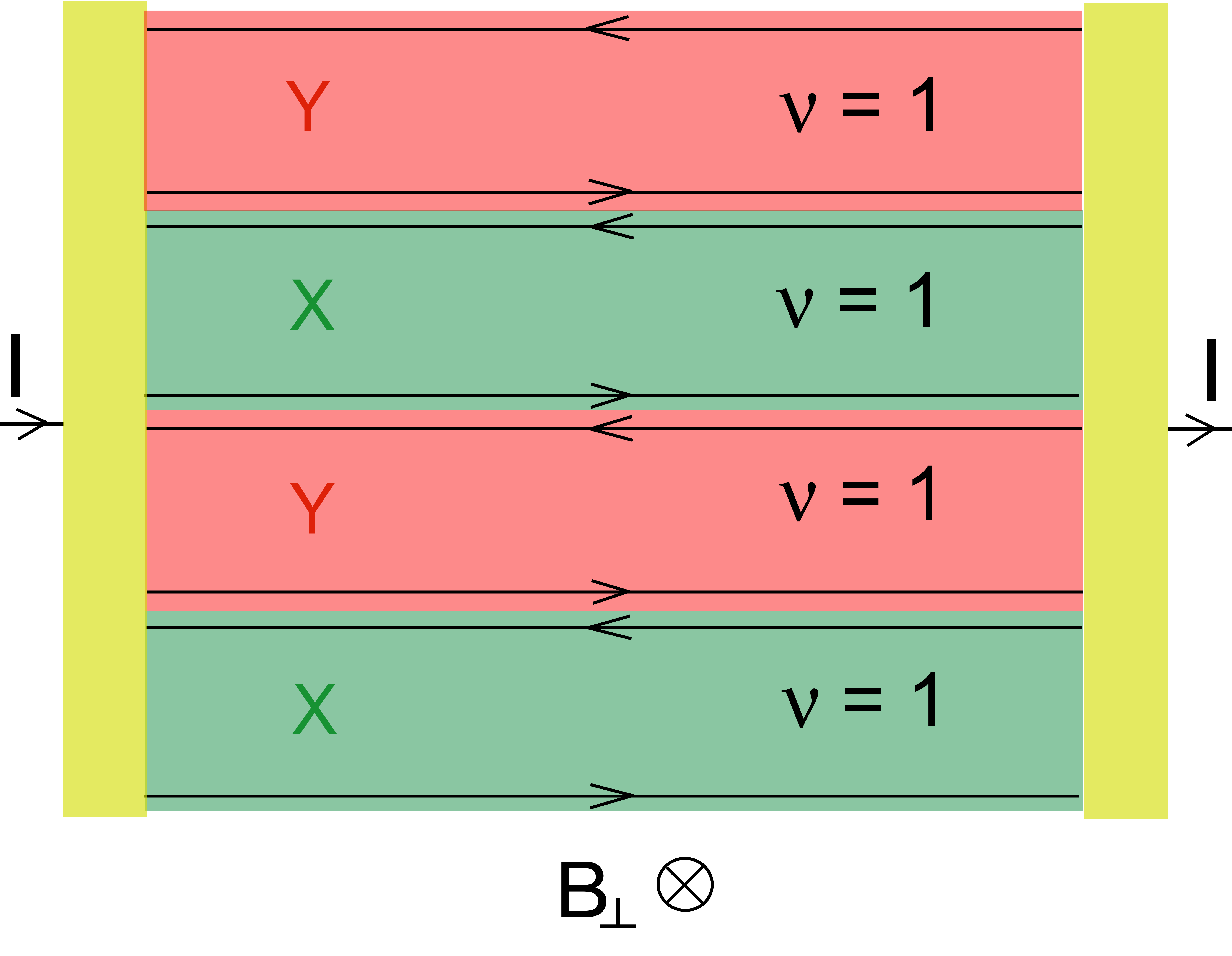}
\caption{\label{fig:Fig1}
Schematic of realizing valley helical edge states from a valley superlattice subjected to large perpendicular magnetic fields.}
\end{figure}

In summary, our results demonstrate a novel technique to induce a periodic valley modulation for 2D electrons in a multi-valley system. We remark in closing on a possible extension of our work to realize a device which could support one-dimensional, helical edge modes (Fig. 3) and therefore be of potential use in exploring Majorana fermion physics. Note that our choice of negative electron-beam resist as the surface grating yields a valley modulation of $\simeq20\%$ in our samples (see Supplemental Material \cite{Supplemental}). This modulation may be enhanced to essentially $100\%$ by using a grating material, e.g. Ti, whose thermal contraction coefficient is significantly different from that of GaAs/AlAs \cite{Footnote1}, meaning that the 2D electrons in alternating strips would occupy either X or Y valleys. As illustrated in Fig. 3, when current is passed parallel to the \textit{valley-polarized} strips at the $\nu=1$ integer quantum Hall state \cite{Klitzing.RMP.1986}, the large $B_{\perp}$-induced one-dimensional edge states should be confined within each strip. This paves the way to realizing counter-propagating, edge channels of different valleys (X and Y) at the boundary between two strips.

The valley-polarized edge modes depicted in Fig. 3 mimic the spin-polarized one-dimensional helical channels where electron's spin is locked to its momentum. Recent years have seen a dramatic surge in the studies of spin helical edge modes, with a particular emphasis on coupling with s-wave superconductors to engineer Majorana fermions \cite{Lutchyn.PRL.2010,Oreg.PRL.2010}. The accessibility to Majorana modes, combined with their potential as the building blocks for fault tolerant quantum computation \cite{DasSarma.npj.2015,Karzig.PRB.2017}, has made the host materials of helical conductors desired systems for future quantum devices. Of particular interest are quantum Hall based systems that provide a robust platform for spin helical edge conduction \cite{Young.Nat.2014,JDSY.NatNano.2017,Ronen.arxiv.2017}. Our proposed valley helical edge modes in the quantum Hall regime (Fig. 3) extend beyond such helical channels of the spin-variety. Although it remains to be seen whether valley, often considered as a pseudo-spin \cite{Shayegan.Physica.2006,Gunawan.PRL.2006,Gunawan.PRB.2006,Shkolnikov.PRL.2004,Shkolnikov.PRL.2005,Gunawan.NatPhy.2007}, can lead to similar Majorana physics as the spin degree of freedom, our results certainly enrich the potential of multi-valley systems for device application.

\begin{acknowledgments}
We acknowledge support through the NSF (Grants ECCS 1508925 and DMR 1709076) for measurements. For sample fabrication and characterization, we also acknowledge the NSF (Grant MRSEC DMR 1420541), the Gordon and Betty Moore Foundation (Grant GBMF4420), and the DOE BES
(DEFG02-00-ER45841). 
\end{acknowledgments}

\end{document}